\documentclass[english,aps,article,showpacs]{revtex4}
\usepackage[T1]{fontenc}
\usepackage[latin1]{inputenc}
\usepackage{babel}
\usepackage{graphics}
\usepackage{setspace}
\makeatletter

\providecommand{\LyX}{L\kern-.1667em\lower.25em\hbox{Y}\kern-.125emX\@}

\usepackage[T1]{fontenc}
\usepackage[latin1]{inputenc}
\usepackage{babel}
\usepackage{graphics}

\makeatletter

\usepackage[T1]{fontenc}
\usepackage[latin1]{inputenc}
\usepackage{babel}

\makeatletter

\usepackage[T1]{fontenc}
\usepackage[latin1]{inputenc}
\usepackage{babel}



\makeatother

\makeatother

\begin{document}

\title{Two early-stage inverse power-law dynamics in nonlinear complex systems far-from-equilibrium}

\author{Somnath Bhattacharya} 

       \email{somnath.bhattacharya@saha.ac.in}

\author{Asok K. Sen\footnote{Corresponding author. Electronic address:
asokk.sen@saha.ac.in}}

\affiliation{TCMP Division, Saha Institute of Nuclear Physics, 1/AF Bidhan Nagar, Kolkata 700 064, India}

\vskip 1.0cm


\begin{abstract}
We consider the dynamics of the charge carriers in a {\it tunneling-enhanced
percolation} network, named as a Random Resistor cum Tunneling-bond Network
(RRTN), where we allow tunneling in the gap between two randomly thrown
nearest neighbour metallic bonds only.  Our earlier studies
involved the dc and ac {\it nonlinear response}, the percolative aspects,
dielectric breakdown, low-temperature variable range hopping (VRH)
conduction, etc. in the RRTN.  Here we study the non-equilibrium dynamics
of the carriers.  With two far-from-equilibrium, initial inverse power-law
relaxations extending over several decades, the dynamics has a lot of
similarities with a wide variety of naturally occuring avalanche-like, run-away 
phenomena in {\it driven}, disordered systems with {\it statistically correlated}
randomness.  In the power-law regime, the RRTN violates the Boltzmann's 
(or, Debye) relaxation time approximation strongly.  Beyond this regime, the 
response relaxes exponentially fast (acquires one time-scale) to a steady-state,
and thus the relaxation approximation becomes exact.

\end{abstract}

\maketitle

\section{Introduction}

Slow relaxation phenomena in random composite materials made of components
having widely different generalized susceptibilities ({\it e.g.}, permeability,
dielectric constant, electrical/ thermal conductivity, viscosity, elastic
modulii etc.), continue to remain intriguing and hence a topic of
intense research.  The susceptibility, being a measure of the {\it response}
of a system to an appropriate external perturbation or a {\it driving}
force, is typically assumed to be {\it linear} under a vanishingly small
external force.  But, many natural systems/ phenomena do not
show any measurable response until an appropriate driving force exceeds a
measurable {\it threshold}.  For example, a rigid body on a rough surface
does not move until the driving force exceeds a finite frictional force.
More intriguingly, it is almost a law of nature that in such {\it driven},
{\it macroscopic} systems, the response characteristic is {\it nonlinear} \cite
{kardar} with a concommitant critical behaviour at the threshold.  Besides
the case of frictional motion, nonlinear response above a threshold is also
found in pinned charge-density-wave systems, disordered granular
superconductors \cite{doniach}, etc.

In studying the dynamical behaviour of a material, one usually measures its
appropriate response property, say, $\phi(t)$, as a function of time $t$,
from a non-equilibrium to an equilibrium (for a
closed system) or to a steady (for an open system) state.  In general,
this relaxation is classified into two groups: (i) a purely Debye type
with an exponential relaxation function, $\phi(t)= {\rm exp}(-t/\tau)$,
$\tau$ being a characteristic time-scale, called the {\it relaxation
time}; or (ii) a
non-Debye type where $\phi(t)$ is either a linear superposition of
exponential functions, or a sub-exponential function (as in some glassy
systems), with {\it multiple} relaxation times.  In more intriguing
cases, this non-Debye relaxation may even be a power-law or a logarithmic
function, without any time-scale, or one with $\tau \rightarrow \infty$.
The appearence of such a
{\it scale-free, slow} dynamics with one or more power-laws from the early
stages of evolution, is what concerns us here because of the fundamental
issue of the failure of Boltzmann's {\it relaxation time approximation}
\cite{zim} in driven systems far-from-equilibrium.

\section{Experiments and some related models}

In a lucidly written review Scher, Shlesinger and Bendler \cite{scher}
focus on experimental observations (1970's onwards) and the origin of
two power-law kinetics.  For a few examples, we cite some transient
photocurrent measurements \cite{tie} on $a$-Si:H, $a$-As$_2$Se$_3$ etc.
and a couple of hole transport \cite{hole} data on PVK and Si-MOS
devices, where two consecutive power-law decays, of the forms
$t^{-\alpha}$ and $t^{-\beta}$ ($\alpha,\beta>0$), covering one or
more decades in time each (with a crossover in-between) were observed.
Based on the {\it continuous time random walk} with a long-tailed
power-law probability density function for the {\it random waiting times}
(release time of trapped carriers by tunneling),
Scher {\it et al.} \cite{scher} formulated a theory regarding the above results.
The latter long-tailed power-law function violates the {\it Central Limit
Theorem}, since all of its moments including the first (mean waiting-time)
diverge.  The unifying feature of the above random
walk is the {\it scale-invariance} of the shape of the relaxation current
$I(t)$, if one normalizes the time by a {\it transit time} $t_r$, which
is a sample dependent parameter.  This stochastic theory by Scher {\it et al.}
\cite{scher} explains the results of many early experiments, following
the relation $\alpha + \beta = 2$.

But, there is a huge variety of relatively recent, more intriguing
experiments on {\it soft-condensed} or {\it complex} systems where the
couple of exponents $\alpha$ and $\beta$ do not seem to follow any simple
algebraic relation.  Weron and Jurlewicz note that an experiment on
dielectric relaxation in a system of dipoles \cite{hill} involves the
crossover between different forms of self-similarity.  They argue that
\cite{wj} a couple of power-law decays appear due to the coupling of
micro-clusters of dipoles with a distribution of $\tau$'s ({\it i.e.}, multiple
time-scales).  An experiment on fluorescence intermittency of {\it single}
ZnS overcoated CdSe quantum dots \cite{kuno}, the distribution of
{\it on} and {\it off} times ({\it blinking kinetics}) is reported to
follow a single power-law
\footnote {Our analysis suggests that two inverse power-law kinetics are
present in this experiment \cite{kuno}, as well.}. 
In a biological system of photo-dissociated heme-proteins, the rebinding
of the ligands of iron (i.e., the CO and the O$_2$ molecules) is observed
to follow an inverse power-law dynamics \cite{pf}.  In a theoretical study
of the same work, Tsallis {\it et al.} \cite{tsal} claim two inverse power-law
regimes and demonstrated its possible connection with {\it nonextensive
thermo-statistics} (entropy).  Naudts and Czachor \cite{naud} analyzed
the data of many experiments including those of \cite{kuno,pf}, and
maintain that these two power-law decays result from some choice of 
parameters of their probability density function.  All of the above
theories lead to the result that $\alpha < \beta$, if two power-law
relaxations exist.  In a different biological system, the dynamics of
Ca channels in living cells \cite{bez}, the distribution of the
survival-times of the channels has been studied.  A stochastic dynamical
model, with one dimensional geometry, was proposed specifically to explain
only the later power-law \cite{bar} dynamics
\footnote {Two inverse power-law kinetics are clearly present in the
Fig.2, a typical result, of this 1-d model \cite{bar}.}.

Two power-law growths are also claimed to have been found, using atomic force 
microscopy (AFM), in the early dynamics of sputtered Ag particles on Si(0~0~1)
substrate \cite{sangam}.  These authors relate the two different growth-dynamics
to two competing structural rearrangements at two different length-scales.
In the case of the growth of a large single DLA (diffusion limited aggregation)
cluster, using upto 10$^8$ particles in a computer experiment, two power-laws
seem to dictate the growth process as a function \cite{mandel} of a time-like
entity.  A 
similar theme has been reflected in various methods ({\it e.g.}, the sol-gel method) of
 synthesis of nanomaterials \cite{edcam}.  This involves a slow relaxation of
restructuring many local clusters by crossing the local barriers due to a
{\it caging effect} (typical of liquid-like, amorphous or dense granular
materials) followed by (or, may be, added to) another slow relaxation of
global restructuring of the local clusters themselves.  Even in some relatively
 recent studies of cellular automata models of earthquakes \cite{naka}, one
power-law dynamics seems to result under certain choice of parameters.  

\begin{figure}
%\onefigure{conf.ps}
\resizebox*{7cm}{7cm}{\rotatebox{270}{\includegraphics{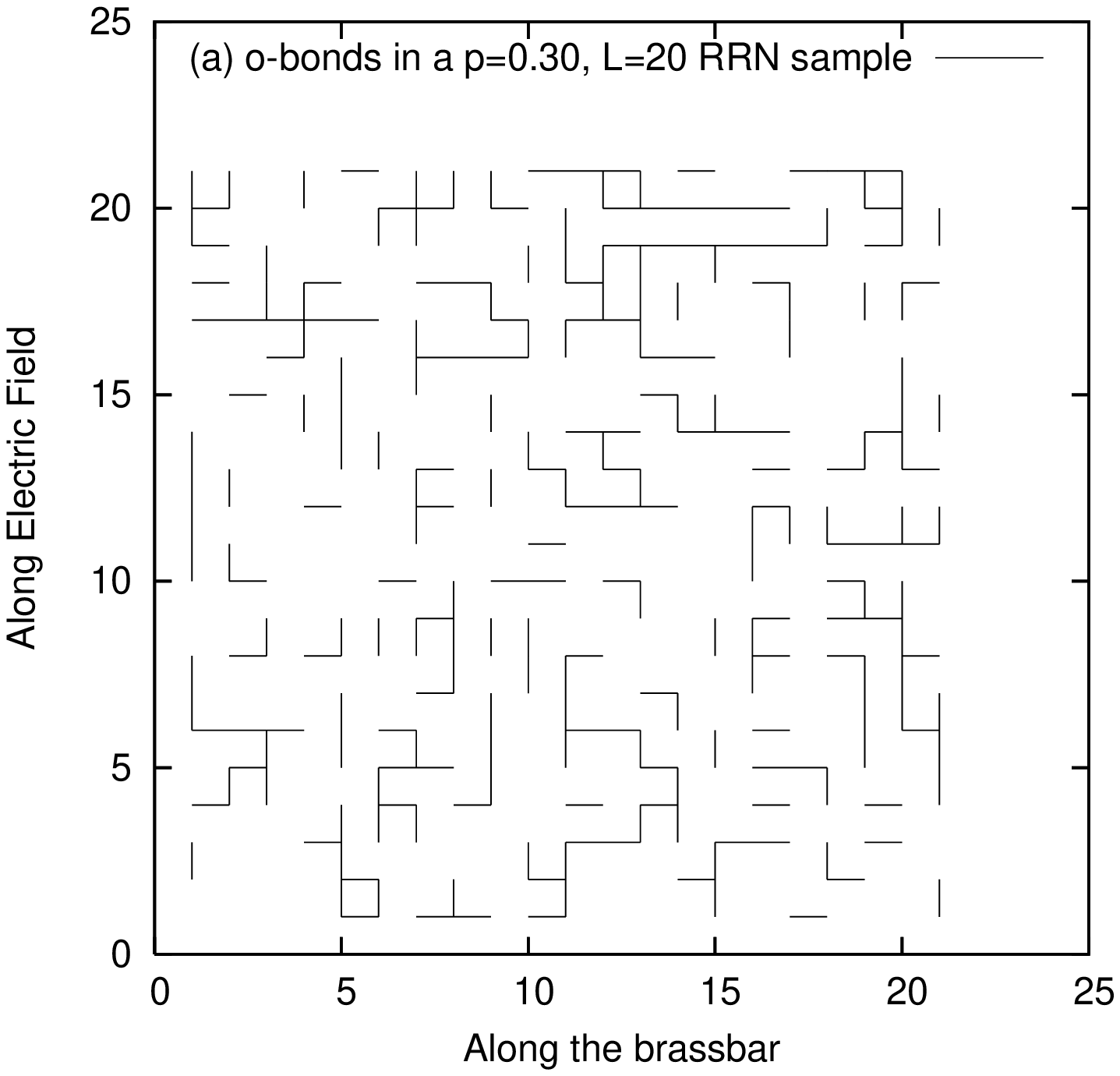}}}
\resizebox*{7cm}{7cm}{\rotatebox{270}{\includegraphics{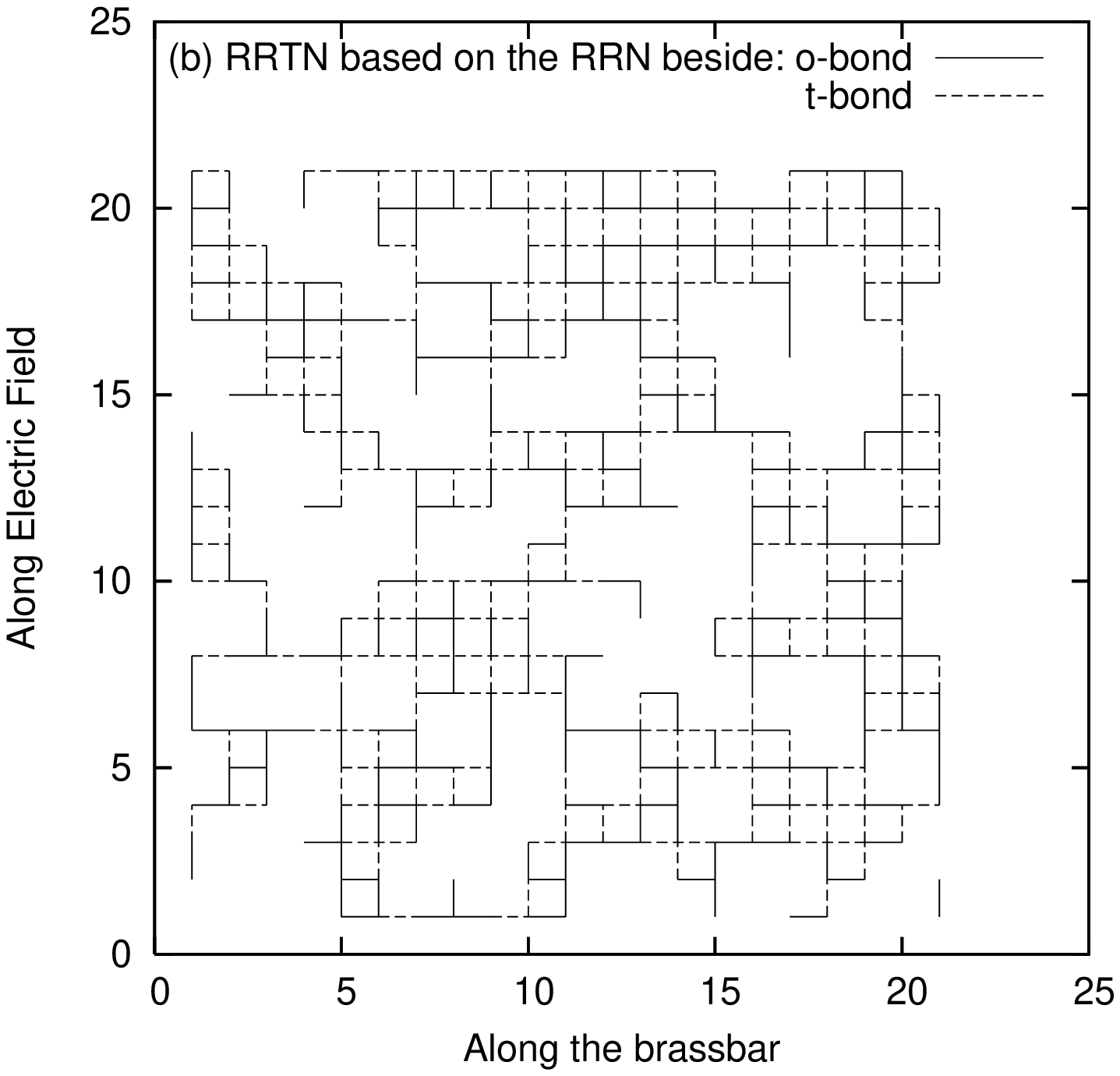}}}
\caption{(a) A typical configuration of random metallic o-bonds in a $L=20$,
$p=0.3$ RRN sample on a square lattice.  The insulating bonds are kept blank.
This RRN configuration, being below its threshold of \(p_c=0.5\), is not
percolating (an $insulating$ RRN); (b) In the ensuing {\it maximal} (with
fully statistically correlated tunneling t-bonds) RRTN configuration, the
t-bonds (dashed lines) are placed between any two nearest neighbour o-bonds.
They get activated when the potential differences across them exceed a
microscopic voltage threshold, \(v_g\).  With all the allowed t-bonds active,
this maximal RRTN percolates (is a $'metallic'$ RRTN).}
\label{f.1}
\end{figure}

\section{A tunneling percolation model}

To study the basic physics behind the transport properties within a
minimal model, Sen and co-workers \cite{dcac,pre1} constructed a lattice
based bond percolation model.  In this model, in addition to the randomly
placed ohmic bonds (linear response) in the Random Resistor Network (RRN),
{\it tunneling} bonds (t-bond) were introduced between two nearest
neighbour ($nn$) ohmic bonds (o-bonds) separated by one insulator only.
This tunneling is actually {\it semi-quantum} or semi-classical in the
sense that no quantum mechanical phase-information of the charge carriers
appears in the model.  In the steady-state, a t-bond is insulating if the
microscopic voltage difference across it ($v$) is less than a fixed
voltage threshold ($v_g$, identical for all the t-bonds) and conducts
linearly a current if $v \ge v_g$.  Indeed, the phenomenological
parameters $p$ and $v_g$ ensure that both the {\it disorder} and the
{\it coulomb interaction} (nonlinear response \cite{mehran} as an
outcome) is in-built in this model.

Historically \cite{dcac}, we call this tunneling-enhanced percolation model
as the {\it Random Resistor cum Tunneling Network} (RRTN).  The appearence
of the t-bonds in this {\it perfectly correlated, i.e., deterministic}
fashion is the origin of a very low percolation threshold, in the
RRTN model.  For example, in the fig.\ref{f.1}(a), the RRN with $p$=0.3,
does not percolate.  In this respect, one may note that the percolation
threshold for the RRN (square lattice) is $p_c$ = 0.50.  But, the {\it
maximal} RRTN (where all the possible t-bonds, both along and across the
field, are active, i.e., conducts a maximum possible current in the steady
state) generated from that particular RRN, percolates as shown in the
fig.\ref{f.1}(b).  Standard calculations with finite-size scaling analysis,
establishes a new percolation threshold of $p_{ct}$ = 0.18 \cite{pre1},
for the maximal RRTN.

Further, because of the finite threshold of the t-bonds, the bulk dc
conductance $G(V)$ (defined either as $\frac{I}{V}$ or as $\frac{dI}{dV}$)
is a strongly {\it nonlinear} curve, as stated earlier.  At a sufficiently
low voltage (below $v_g$), no t-bond is active, and hence the effective
bulk conductance is only due to the percolating ohmic backbone.  So the
$G(V)$ vs $V$ graph is parallel to the $V$-axis (falls on the $V$-axis if
the ohmic backbone is non-percolating).  This region is termed as the lower
linear/ohmic regime.  Thereafter, under further increment of the external
voltage $V$, yet more microscopic t-bonds cross the threshold ($v_g$),
i.e., become active, and give rise to extra parallel paths in the network.
At an very high voltage \cite{foot2} (asymptotically infinite),
all the possible t-bonds become active (i.e., creates a
maximal RRTN) and hence the conductance saturates to its largest possible
value (for that particular configuration).  Obviously, this region belongs
to the upper linear/ohmic regime.  Other than the success of this discrete
model in understanding various experiments on an ultra-low percolation
threshold, and various aspects of nonlinear dc and ac responses, it
has also been quite useful for understanding some very unusual aspects
observed in low-temperature variable range hopping (VRH) conduction
\cite{vrh}, and some interesting aspects of breakdown phenomena
\cite{break}.  Thus, even though, time enters in an implicit fashion in
some of the above studies, an explicit characterization of the relaxation
dynamics in the RRTN model in the perspective of various experiments and
models (as described above) was considered necessary.

\section{Non-equilibrium dynamics in the RRTN model}

In this paper, we study the current relaxation dynamics towards a
steady-state in the RRTN model.  Away from equilibrium, a t-bond with
a microscopic $v < v_g$ behaves like a dielectric material between two
metals (o-bonds); and the resulting charging effect gives rise to a {\it
displacement current} ($C\frac{dv(t)}{dt}$, where $C$ is the capacitance).
Thus, a t-bond gives rise to a displacement current if $v(t) < v_g$, and an
`ohmic' current if $v(t) \ge v_g$ \cite{sust}.  For our calculations, we use
the values of the microscopic conductance $g_o=1.0$ (o-bonds), $g_t = 10^{-2}$ 
(t-bonds), $v_g=0.5$, and $C=10^{-5}$ for the t-bonds (in some arbitrary units).

%\label{}
In our numerical study, we apply an uniform electric field across
RRTN's of different system sizes ($L$) and ohmic bond concentrations
($p$).  We study the evolution of the current in a RRTN starting from the
switching on state until it approaches its {\it asymptotic} steady-state.
To do this, we follow basically the current conservation (Kirchhoff's
laws) locally at each node of the lattice.  The aim is to study the
achievement of a {\it global current conservation} as an outcome of the
{\it local current conservation} (hence, the dynamics).  A discrete, {\it
scaled} time unit has been chosen as one complete scan through each site
of the lattice.  This local conservation or the equation of continuity
reads as, 

\begin{equation}
%\begin{displaymath}
\label{eq.1}
\sum i_{nn}(t) = 0, \hskip 2.0cm  {\forall t}.  
\end{equation}
%\end{displaymath}

\noindent
Here the {\it sum} has been taken over currents $i_{nn}(t)$ through
various types of nearest neighbour ($nn$) microscopic bonds
around any node/site of the lattice.  For the case of a square lattice,
one considers the four $nn$'s around a node inside the bulk (three and
two $nn$'s respectively at any boundary or a corner).  If eq.~(\ref{eq.1})
were true simultaneously for each site of the lattice, then the global
conservation (the steady state) for the entire network would automatically
be achieved.  As we need to start with an initial (arbitrary) microscopic
voltage distribution, the eq.~(\ref{eq.1}) would not
hold for all the sites of the lattice.  Some correction term would be
required at each site and this requirement leads to the following time
evolution algorithm which we call as the {\it lattice Kirchhoff's dynamics}:

\begin{equation}
%\begin{displaymath}
\label{eq.2}
{v(j,k,t+1)} = {v(j,k,t)} + \frac{\sum i_{nn}(t)} {\sum g_{nn}},
\end{equation}
%\end{displaymath}

\noindent
where $g_{nn}$ are the various microscopic conductances of the $nn$
bonds around the node $(j,k)$.  Then we numerically solve a set of
coupled {\it difference} equations on the lattice.
The move towards a macroscopic steady-state
implies that the difference of currents through the first and the last layers
tends to zero as a function of time.  In practice, the system is considered
to have reached its steady state when this
difference decreases to a pre-assigned smallness.

\begin{figure}
%\onefigure{dynap.55l60v2.ps}
\resizebox*{7cm}{7cm}{\rotatebox{270}{\includegraphics{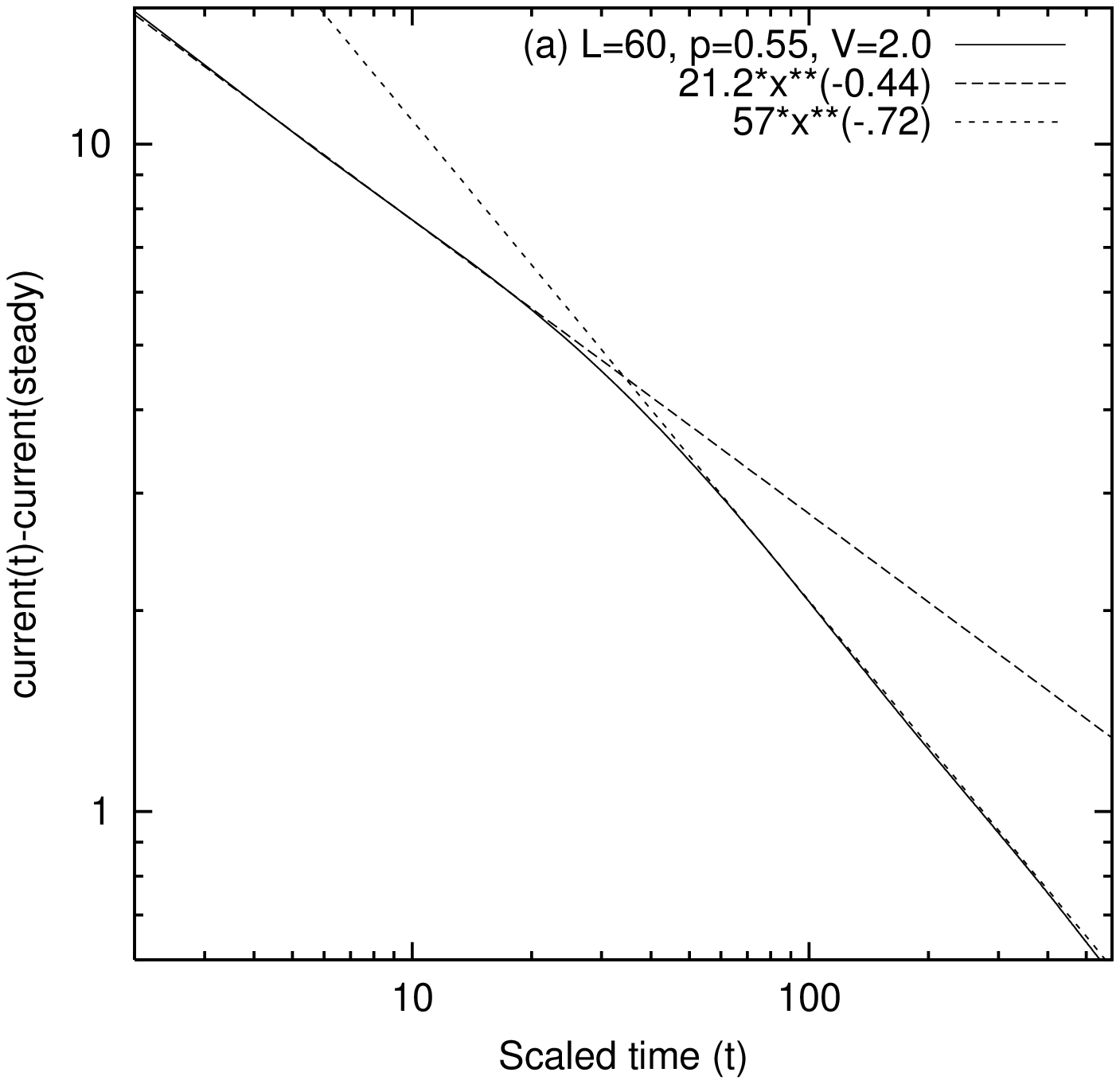}}}
\resizebox*{7cm}{7cm}{\rotatebox{270}{\includegraphics{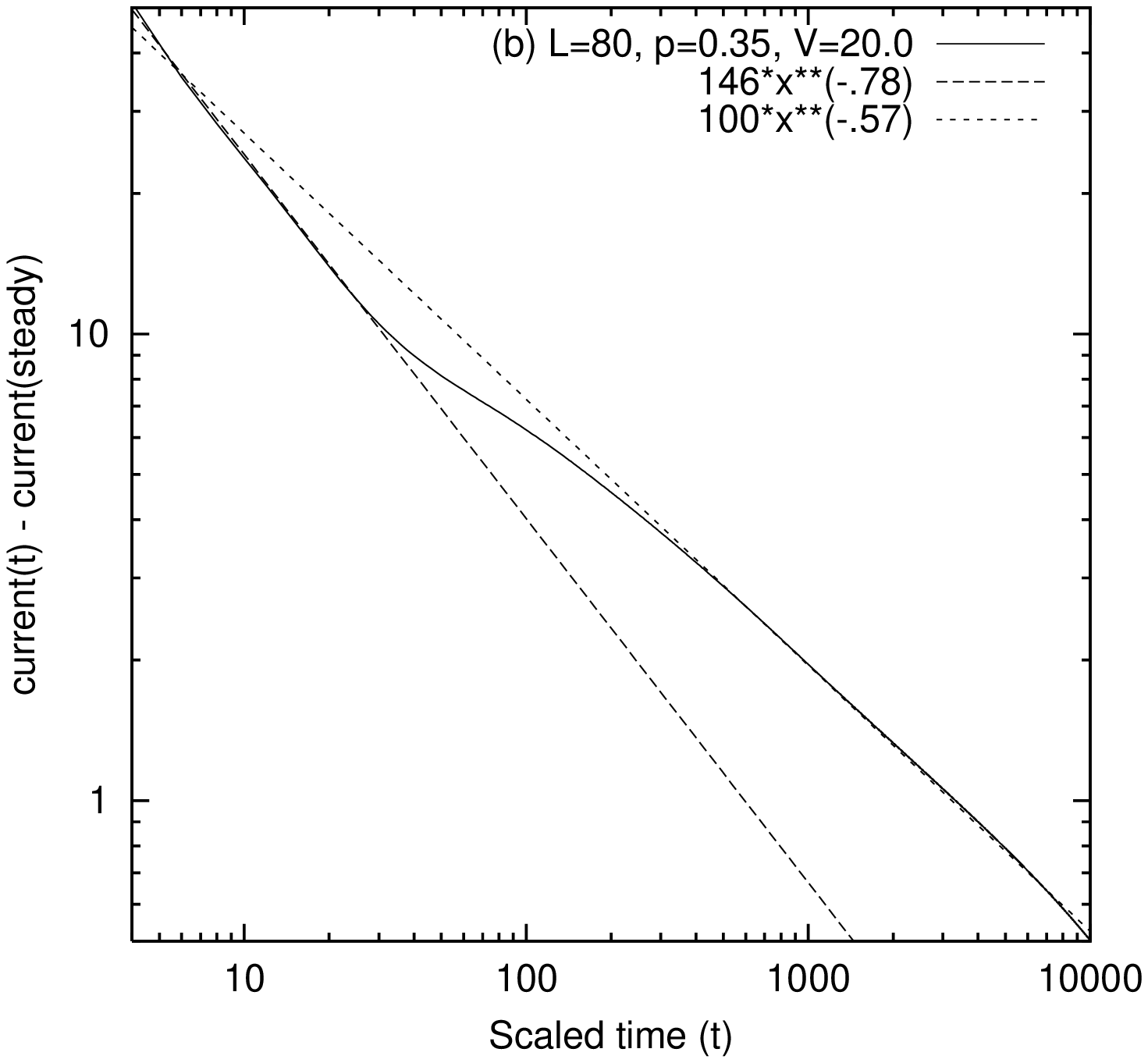}}} 
\caption{Non-equilibrium dynamics in terms of a {\it scaled time} shows two
initial power-law decays (with intermediate crossover) in the current
relaxation for (a) one RRTN sample (\(p=0.55, L=60\)) with the
exponents of \(0.44 \) and \(0.72 \); and (b) another sample (\(p=0.35,
L=80\)) where the exponents are \(0.78\) and \(0.57\) respectively.
The dynamics in the RRTN is always followed by an {\it exponentially}
decaying tail (not shown in the figures here) towards its steady state.}
\label{f.2}
\end{figure}

\section{Results and discussions}

As shown in the figs.~\ref{f.2}, we observe a
non-Debye type current relaxation with two consecutive initial power-laws
(and a crossover in-between), each of them spanning at least a decade.  This
asymptotic steady state current (whether insulating or conducting) for any
randomly chosen RRTN, is found to be very robust against any initiallly
chosen voltage distribution on the lattice.  To analyse the current
evolution of an insulating or a conducting sample on the same footing,
we subtract the corresponding steady current, $I(t~\rightarrow~\infty)$,
from the evolving current $I(t)$ at the time $t$.  Our observation on the
transient current response, indicates clearly the existence of a 
couple of initial power-laws, whose exponents differ significantly for
systems with different configurations, $p$, $L$ and external voltage $V$.
So, we choose to work with one sample at a time and analyze its results
within our numerical accuracy.

For example, in the fig.~\ref{f.2}(a), we show the dynamics for a sample
with $L=60, p=0.55, V=2.0$ and in the fig~\ref{f.2}(b) another with
$L=80, p=0.35, V=20.0$.  The first figure ({\it i.e.}, fig.~\ref{f.2}(a))
represents the class of relaxation, where the second exponent ($0.72$)
is larger than the first ($0.44$) (the only class reported in our previous
work at a particular $p=p_c$ \cite{sust}).  There are quite a few
theoretical works
\cite{scher,wj,tsal} in this regard.  Some other experiments \cite{bez,bar,
sangam} find a second exponent smaller than the first.  Since our earlier
report \cite{sust}, we have been able to reproduce this other class of
relaxation ({\it e.g.}, fig.~\ref{f.2}(b) with the exponents $0.78$, $0.57$), as
well, within the context of our RRTN.  In special cases, we do find only
one power-law relaxation, which may be considered to be the borderline
between the above two, or as the merging of the two power-law exponents.
Further, we do not find any particular relation between the exponents.
Next, the existence or the lack of any asymptotic exponential kinetics is not
explicitly stated in most of them.  Indeed, in some of the theoretical
studies ({\it e.g.} in \cite{scher,wj,mandel}), the second power-law persists
upto asymptotically infinite times.  This trend can not describe the
possiblity for these systems to reach an appropriate steady state.
In contrast, beyond the power-law relaxations (whether one or two), our
model acquires
a relaxation time $\tau$ and the system follows a fast exponential dynamics
as a precursor of a final steady state which is effectively diffusive
(or ohmic).  Obviously, the power-law relaxations at times $t < \tau$
imply a strong deviation from the Boltzmann's {\it relaxation time
approximation} (i.e., strongly non-Debye type).

As far as the origin of the two power-law dynamics are concerned, we have
already outlined the main content of some of them \cite{scher,wj,tsal,bar,
naud,naka} in the section on experiments.  In most of them they occur due
to local structural rearrangements preceding the final global structural
rearrangements.  In our case, the basic structure in a particular sample
is created once for all, and it is the fields across the bonds which
keep changing in such a fashion that the local conservation (Kirchhoff's
laws at each node) dominates the first power-law regime and the
global current conservation is dominated in the latter power-law regime
(of course, there are the required structural rearrangements of the
active t-bonds).  So, it is interesting that these two very different
mechanisms give rise to a qualitatively identical outcome.  Further,
since the power-law dynamics occurs even for $p$'s away from $p_{c}$ or
$p_{ct}$, it clearly demonstrates that they are not organized by any type
of criticality.  Finally, as discussed above, while most of the other
theoretical works,
are destined to get only one class of two-power-law relaxation behaviour
(namely, $\alpha < \beta$), the RRTN dynamics has the ability to capture
both of the classes for different sets of parameters.

\acknowledgments
Sincere thanks are due for warm supports to the organizers of many
International Conferences where earlier results on relaxation in the RRTN were
presented.  Foremost among them are Profs. P. Sheng and P.M. Hui for
one named ETOPIM5 (AKS in 1999) was held at Hong Kong.  Two more on Nonlinear
Dynamics and Earthquake (SB in SMR1322 in 2001 and SMR1519 in 2003) were
held at the AS-ICTP, Trieste, Italy.  The last one named ICAMMP2002 (AKS
and SB in 2003) were held at the SUST, Sylhet, Bangladesh.  AKS also
acknowledges fruitful discussions with Prof. H. Levine in 2003 on their
Ref.\cite{bar} and the support (a Visiting Professor in 2004) plus
discussions with Prof. A. Hansen of the NTNU, Trondheim, Norway.


\begin{thebibliography}{10}
\bibitem{foot2}
    Note that our aim in the paper on the response characteristics in the
    RRTN is to study the nonlinear effects due to tunneling only.  So we
    do not consider any Joule heating in our system.
\end{thebibliography}
\end{document}